# New applications of the H-reversal trajectory using solar sails

Xiangyuan Zeng, Hexi Baoyin, Junfeng Li, and Shengping Gong

School of Aerospace, Tsinghua University, Beijing 100084, China;

Corresponding Email: cengxy08@mails.tsinghua.edu.cn

**Abstract:** Advanced solar sailing has been an increasingly attractive propulsion system for highly non-Keplerian orbits. Three new applications of the orbital angular momentum reversal (H-reversal) trajectories using solar sails are presented in this paper: space observation, heliocentric orbit transfer, and collision orbits with asteroids. A theoretical proof for the existence of double H-reversal trajectories (referred to as 'H2RTs') is given, and the characteristics of the H2RTs are introduced before the discussion of the mission applications. A new family of H2RTs was obtained using a 3D dynamic model of the two-body frame. In a time-optimal control model, the minimum period H2RTs both inside and outside the ecliptic plane were examined using an ideal solar sail. Due to the quasi-heliostationary property at its two symmetrical aphelia, the H2RTs were deemed suitable for space observation. For the second application, the heliocentric transfer orbit was able to function as the time-optimal H-reversal trajectory, as its perihelion velocity is circular or elliptic velocity. Such a transfer orbit can place the sailcraft into a clockwise orbit in the ecliptic plane, with a high inclination or displacement above or below the Sun. The third application of the H-reversal trajectory was simulated impacting an asteroid passing near Earth in a head-on collision. The collision point can be designed through selecting different perihelia or different launch windows. Sample orbits of each application were presented through numerical simulation. The results can serve as a reference for theoretical research and engineering design.

## 1. INTRODUCTION

Solar sailing has long been considered to be an attractive propulsion system without the use of propellant mass. Since NASA adopted solar sailing as a possibility for a Halley's comet rendezvous mission[1] (Friedman et al., 1978), a lot of research has been conducted using two-body dynamics with low-performance solar sails (Sauer, 1976). As the world's first spacecraft to use solar sailing as the main propulsion, IKAROS[2] (Interplanetary Kite-craft Accelerated by Radiation Of the Sun), launched by Japan Aerospace Exploration Agency (JAXA), successfully demonstrated solar sailing technology in interplanetary space (Mori et al., 2010). In a sense, the IKAROS mission is analogous to the celebrated first flight by the Wright brothers. However, when designing future missions, the engineer's imagination is limited by inevitable real-world constraints. While low-performance solar sails can complete interplanetary transfer using heliocentric spiral trajectories (Macdonald et al., 2007), high-performance solar sails can enable exotic applications using non-Keplerian orbits (McInnes, 1999). Although these trajectories require advanced sail designs and materials, it is still necessary to investigate the exotic trajectories with current sail concepts.

---

[1] http://www.planetary.org/solarsailcd/friedman.htm
[2] http://www.jspec.jaxa.jp/e/activity/ikaros.html

After the introduction of the H-reversal trajectory about ten years ago, due to the high performance of solar sails, subsequent research only focused on the solar escape trajectory. Vulpetti was the first to address the H-reversal trajectory in 1996, and he investigated both 2D and 3D H-reversal trajectories, including the dynamics and applications for interstellar missions (Vulpetti, 1996; 1997; 1998). The sailcraft in H-reversal mode can achieve a high speed to escape the solar system with a single solar photonic assist. However, the sailcraft in H-reversal mode presented by Vulpetti would always move in fixed attitude angles during its maneuvers. Recently, some new features and applications of the H-reversal trajectory have been found. Initial research concerning the H-reversal trajectory attempted to find an accurate feasible region of the sail lightness number for the fixed cone angle trajectory. A new type of double H-reversal periodic orbit was discovered accidentally within fixed cone angle. The trajectory was also addressed by Mengali (2010), referred to as the "H2-reversal trajectory" (H2RT).

This type of orbit is the subject of this paper, and some new features of the orbit are revealed. Based on these new features of the H-reversal trajectory, three main mission applications are discussed in detail. A theoretical proof for the existence of both 2D and 3D H2RTs is given, and the characteristics of 2D H2RTs are briefly introduced. Actually, the so-called '2D H2RT' located in the ecliptic plane was obtained without simplification of the 3D dynamic model. The minimum period H2RTs were simulated in a time-optimal control model with an ideal sail using an indirect approach (Mischa et al., 2005). Certain special H2RTs, such as the 'Self-loop orbit', are also discussed.

The first application of the H-reversal trajectory is space observation with the H2RT. The near regions of the H2RT's two symmetric aphelia are similar to those of the quasi-heliostationary orbit (Mengali et al., 2007), which is very suitable for space observation. The aphelia of H2RTs can range from 1AU (Astronomical Unit) to extremely far away from the Sun. Among the family of H2RTs in the ecliptic plane, the space from the Earth's orbit to Saturn's orbit or further can be detected theoretically, while space up or down the ecliptic plane can be observed using 3D H2RTs. Considering the data transmission between the sailcraft and the Earth, the method for required period H2RTs is presented for use in future mission design.

The second application is the heliocentric transfer orbit. With an H-reversal trajectory, the sailcraft can be transferred to three kinds of heliocentric orbits. All of them are clockwise. In contrast to Vulpetti's hyperbolic velocity at the perihelion, the perihelion velocity of H2RTs is less than the escape speed. Therefore, if there is no longer sail force after the perihelion time, the sailcraft will orbit the Sun in a circular or elliptic orbit. With a half trajectory of 2D H2RTs, the final orbit of the sailcraft will be in the ecliptic plane with a predetermined perihelion or aphelion. For the 3D case, the sailcraft can be transferred to a Sun-centered displaced orbit or a high inclination orbit. It is easy to achieve the altitude of the displaced orbit by varying the H2RT. The orbit inclination can be designed in analytical form.

The third application concerns collisions with asteroids near Earth. To eliminate threats from asteroids passing close to the Earth, one of the three main strategies is collision with the asteroids (Jonathan, 2003). For such a strategy, the trajectory in H-reversal mode is a preferable option because it can produce high crash energy in a head-on impact. For a collision, the trajectory can be designed to make the sailcraft obtain a hyperbolic velocity at its perihelion.

There are still many issues to be addressed regarding the H-reversal trajectory. The minimum lightness number of the example in this paper is 0.5 corresponding to 2.965mm/s$^2$ at 1AU. In future research, the actual minimum sail lightness number required to achieve an H-reversal trajectory should be investigated in an optimal framework. Also, new H2RTs will need to be revealed to give a more general view of their application. Although the required high-performance solar sails are beyond the scope of currently available technology, the new applications of the H-reversal trajectory can still give inspiration to engineers.

## 2. THE H2-REVERSAL TRAJECTORY

### 2.1 Dynamics of the H2-reversal Trajectory

Consider that a solar sail starts from an initial circular orbit of radius $r_0$ to produce the H2-reversal trajectory (H2RT) in the ecliptic plane. An ideal plane solar sail is assumed throughout this paper. The lightness number of the sail is used to describe the solar radiation pressure acceleration that can be expressed as

$$\boldsymbol{f} = \beta \frac{\mu}{r^2} \cos^2 \alpha \, \boldsymbol{n} \tag{1}$$

where $\beta$ is the sail lightness number, $r$ is the distance from the Sun to the sail, $\mu$ is the solar gravitational constant, $\boldsymbol{n}$ is the unit vector directed normal to the sail surface, and $\alpha$ is the cone angle between the sail normal direction and the sunlight. In the two-body frame, all kinds of perturbation forces are not considered and only the solar gravity and solar radiation pressure forces exert on the solar sail. The Cartesian inertial frame *oxyz* (HIF: Heliocentric Inertial Frame) and orbital frame *rth* (HOF: Heliocentric Orbital Frame) (shown in Fig.1) are introduced to illustrate the H2RT. A system of non-dimensional units is applied for convenience. The distance unit is taken as 1 AU, while the time unit is chosen so that the solar gravitational parameter is unitary. Therefore, the dynamical equation of motion in the ecliptic inertial frame can be given by

$$\begin{cases} \dot{\boldsymbol{R}} = \boldsymbol{V} \\ \dot{\boldsymbol{V}} = -\frac{1}{R^3}\boldsymbol{R} + \beta \frac{1}{R^4}(\boldsymbol{R} \cdot \boldsymbol{n})^2 \boldsymbol{n} \end{cases} \tag{2}$$

where $\boldsymbol{R} = \|\boldsymbol{R}\| \boldsymbol{r} = r\boldsymbol{r}$ and $\boldsymbol{V}$ are the position and velocity vectors of the sailcraft, respectively. The parameter $\boldsymbol{r}$ is the unit vector of Sun-sail direction. As shown in Fig.1, the sail acceleration vectors in the frame *rth* can be written as

$$\begin{aligned} f_r &= \beta \cos^3 \alpha / r^2 \\ f_t &= \beta \cos^2 \alpha \sin \alpha \cos \delta / r^2 \\ f_h &= \beta \cos^2 \alpha \sin \alpha \sin \delta / r^2 \end{aligned} \tag{3}$$

where the cone angle $\alpha \in [0, \pi/2]$ and the clock angle $\delta \in [0, 2\pi)$ are the two attitude angles of the sail

(McInnes 1999).

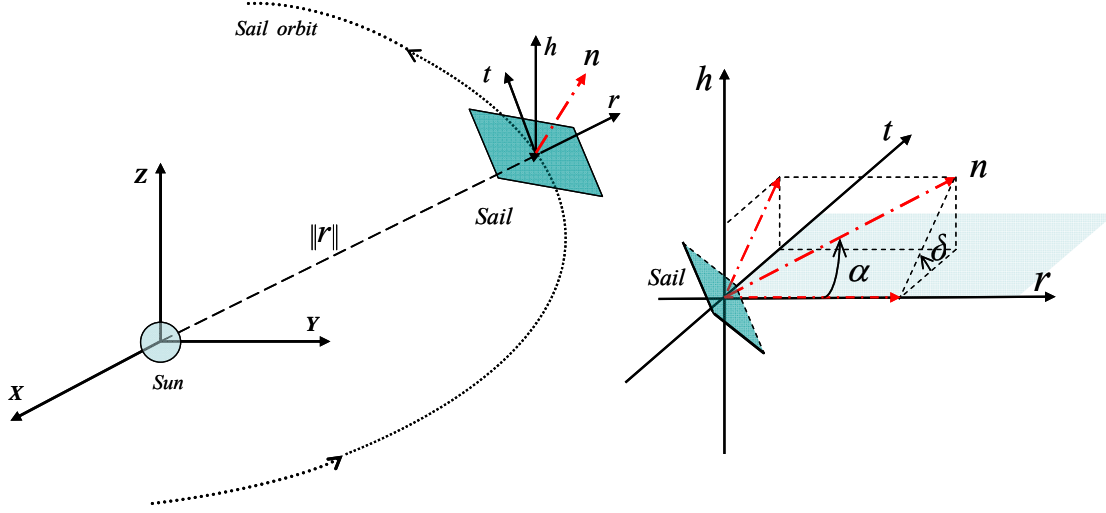

Fig.1　Orientation of the sail cone and clock angles

　　In order to give a clear vision of the H2RT, the 2D case will be introduced briefly at the beginning of this section. Some important features of the H2RT will be illustrated to give a basic understanding of the new non-Keplerian orbits. The 2D equations of motion in the inertial frame can be simplified from Eq.(2) as

$$\ddot{x} = -\frac{x}{r^3} + f_{(\alpha,\beta,r^2)} cos\theta_1$$
$$\ddot{y} = -\frac{y}{r^3} + f_{(\alpha,\beta,r^2)} sin\theta_1$$
(4)

where $\theta_1$ is the angle between the sail normal direction and the inertial x-axis with $\theta_1 \in [0, 2\pi)$. $f_{(\alpha,\beta,r^2)}$ is the magnitude of the solar radiation pressure acceleration which is the same in the frame of *oxyz* and *rth*, expressed as $f_{(\alpha,\beta,r^2)} = \beta cos^2 \alpha / r^2$. In the optimal sail orientation, the parameters $\theta_1$ and $f_{(\alpha,\beta,r^2)}$ are time-varying. For a given trajectory $(t, x(t), y(t), \theta_1)$ at any time $t$, there is guaranteed to be another trajectory $(-t, x(-t), -y(-t), \tilde{\theta}_1)$ which can be seen from Fig.2 with $\tilde{\theta}_1 = -\theta_1$. The equations of motion are invariant under such transformation (Koon et al. 2007; Wilczak et al. 2003) while the value of $f_{(\alpha,\beta,r^2)}$ is the same at time *t* and corresponding time *-t*. As long as the first semi-period parameter $(t, x(t), y(t), \theta_1)$ can be found, the H2RT must exist on the *x*-axis symmetry. In actual, the evolution of the sail normal vector ***n*** is symmetric with *x*-axis to guarantee the symmetry of the H2RT. With the symmetric property, a half trajectory will be enough to analyze the characteristics of the H2RT.

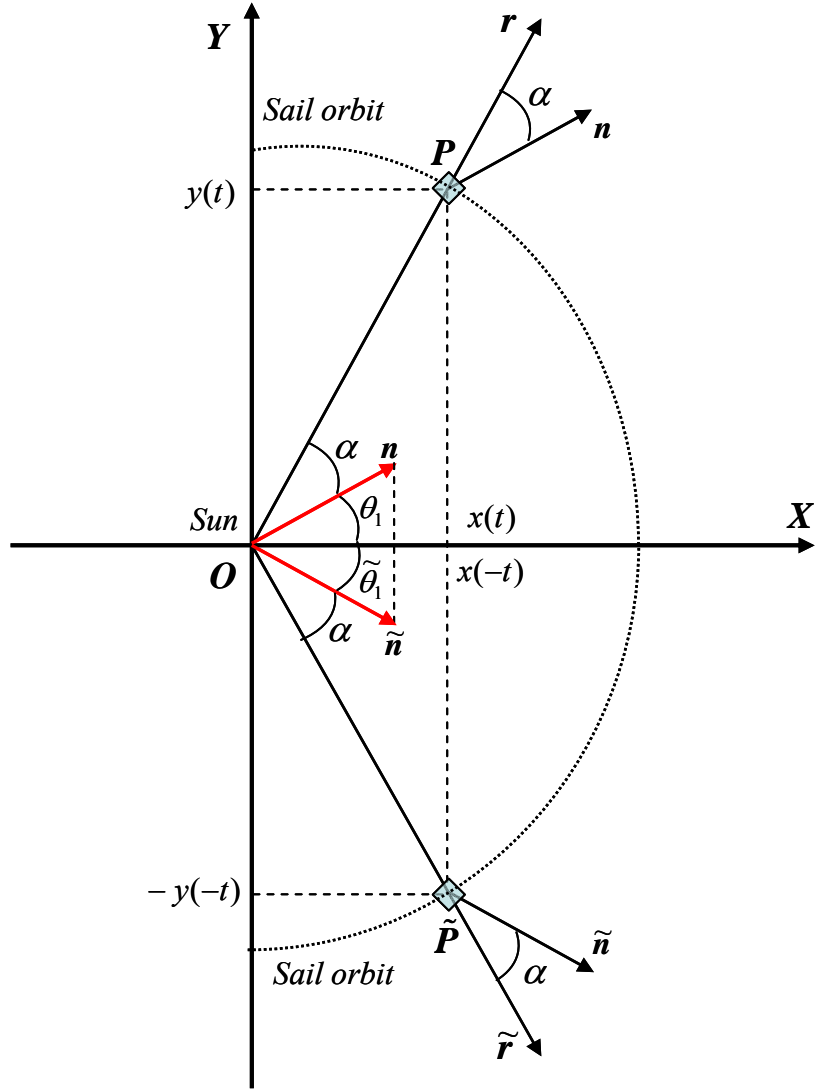

Fig.2　Symmetric structure map for the 2D H2RT

The H2RT is shown in Fig.3 with sail lightness number 0.5 corresponding to 2.965 mm/s² at 1AU. Some important points of the H-reversal trajectory are shown in Fig.3a. The sailcraft starts from default point *A* located in the 1AU parking orbit. After passing its aphelion *B*, it arrives at the zero-H point *C*. Then, it will reach its perihelion *D* and begin the return trip to point *A*. Fig.3b shows the round trip of the H2RT. Although the first semi-trajectory shown in Fig.3a is similar to that found by Vulpetti (1997), the control history of the attitude angles is totally different, which will be discussed later.

Since the 2D H2RT has been proven to be located in the ecliptic plane, is the 3D H2RT symmetrical with a plane as well? For this problem, the 3D dynamic equations in the inertial frame should be given as

$$\ddot{x} = -\frac{x}{r^3} + f_{(\alpha,\beta,r^2)}\cos\theta_1$$

$$\ddot{y} = -\frac{y}{r^3} + f_{(\alpha,\beta,r^2)}\sin\theta_1\cos\theta_2 \quad (5)$$

$$\ddot{z} = -\frac{z}{r^3} + f_{(\alpha,\beta,r^2)}\sin\theta_1\sin\theta_2$$

Similar to the 2D process, for a given trajectory ($t$, $x(t)$, $y(t)$, $z(t)$, $\theta_1$, $\theta_2$) which is the solution of Eq.(5), if ($-t$, $x(-t)$, $-y(-t)$, $z(-t)$, $\tilde{\theta}_1$, $\tilde{\theta}_2$) is also a solution, the 3D H2RT is guaranteed to be symmetrical with plane-$xoz$. If $\tilde{\theta}_1=\theta_1$ ($\theta_1 \in [0, 2\pi)$), Eq.(5a) will be invariant. To ensure the invariability of Eq.(5b), the value of required angle $\tilde{\theta}_2$ should fulfill the expression $\tilde{\theta}_2=\mathrm{mod}(2\pi+\pi-\theta_2, 2\pi)$ since $\theta_2 \in [0, 2\pi)$. With the angles $\tilde{\theta}_1$ and $\tilde{\theta}_2$, Eq.(5c) is also invariant. The above discussion represents a theoretical proof for the existence of both 2D and 3D H2RTs.

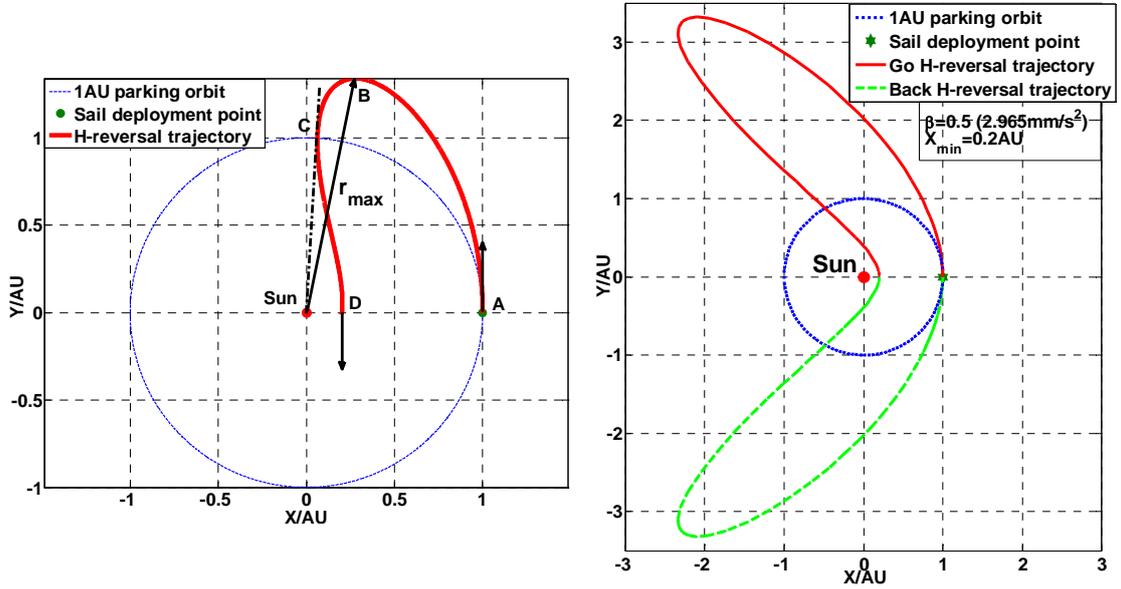

Fig.3    Descriptions of the plane $H^2$-reversal trajectory. Left (a): The H-reversal trajectory. Right (b): $H^2$-reversal trajectory with given sail lightness number 0.5

## 2.2 Features of the H2RT

The time optimal control framework (McInnes 1999) is adopted to simulate the H2RT. Due to axis-symmetrical of the H2RT, only half of the periodic trajectory needs to be calculated during mission design. After the achievement of the first-half-period H2RT, the symmetrical part can be obtained through transformation based on the discussion of Section 2.1. The derivation of the optimal control is briefly outlined to clearly state the problem. The sailcraft starts from the 1 AU circular orbit at zero hyperbolic excess speed ($C_3=0$). Therefore, the initial position and velocity vector can be easily determined. According to the characteristic of the H2RT, the final constraint at perihelion can be expressed as

$$\Psi_f\left[t_f, \boldsymbol{R}(t_f), \boldsymbol{V}(t_f)\right] = \begin{bmatrix} x(t_f)-x_f & y(t_f) & z(t_f)-z_f & V_x(t_f) & V_z(t_f) \end{bmatrix} = 0 \quad (6)$$

where $t_f$ is the time of perihelion passage. The final constraint $z_f = 0$ corresponds to the situation that the perihelion of H2RT locates in the ecliptic plane. While $z_f \neq 0$ the perihelion of H2RT will be outside the ecliptic plane. The object function of minimum period H2RT is given by

$$J = -\lambda_0 \int_0^{t_f} dt \tag{7}$$

where $\lambda_0$ is a positive constant. The Hamilton function of the system is defined as

$$H = -\lambda_0 + \lambda_R(t) \cdot V + \lambda_V(t) \cdot \left[ -\frac{1}{R^3} R + \beta \frac{1}{R^4} (R \cdot n)^2 n \right] \tag{8}$$

where $\lambda_R(t)$ and $\lambda_V(t)$ are the co-states for position and velocity. The velocity co-state variable is also referred to as the primer vector and defines the optimal direction for the solar radiation force vector. The time derivative of the co-states is from the Euler-Lagrange equations

$$\begin{cases} \dot{\lambda}_R = -\frac{\partial H}{\partial R} = \frac{1}{R^3} \lambda_V - \frac{3}{R^5} (R \cdot \lambda_V) R - 2\beta \frac{1}{R^4} (R \cdot n)(n \cdot \lambda_V) \left[ n - \frac{2(R \cdot n) R}{R^2} \right] \\ \dot{\lambda}_V = -\frac{\partial H}{\partial V} = -\lambda_R \end{cases} \tag{9}$$

The optimal sail orientations are obtained by maximizing $H$ at any time expressed as

$$n(t) = \arg\ \max H(t, n, \lambda) \tag{10}$$

Actually, the optimal control of the sail is a locally optimal control with sail orientation defined as

$$n = \frac{\sin(\alpha - \tilde{\alpha})}{\sin \tilde{\alpha}} \frac{R}{\|R\|} + \frac{\sin \alpha}{\sin \tilde{\alpha}} \frac{\lambda_V}{\|\lambda_V\|} \tag{11}$$

where $\tilde{\alpha}$ is the cone angle of the primer vector. The corresponding co-states satisfy the following equation

$$\lambda_V(t_f)_y = 0 \tag{12}$$

The final stationary condition is given by

$$H(t_f) = 0 \tag{13}$$

Since the equations of the co-states are homogeneous, a solution to the equations multiplied by a factor will also be feasible solution. Take the Hamilton function to be scaled to match the transversality conditions, which can be achieved through adjusting the positive constant $\lambda_0$. After normalization the co-states and $\lambda_0$ can be fixed in a unit sphere. The related variables in the normalization at the initial time should satisfy the following equation

$$\sqrt{\lambda_0^2 + \lambda_R^T(t_0) \cdot \lambda_R(t_0) + \lambda_V^T(t_0) \cdot \lambda_V(t_0)} = 1 \tag{14}$$

The concerned free variables are eight, including the perihelion time $t_f$, initial values of the co-states $\lambda_R(t_0)$ and $\lambda_V(t_0)$, and $\lambda_0$. The number of equations with variables constraints, transversality conditions and stationary conditions are also eight, listed as Eq.(6), Eq.(12), Eq.(13) and Eq.(14). There are eight free parameters with eight equations. The main difficulty is to search the appropriate initial values of the co-states to generate the optimal trajectory. An indirect method (Quarta and Mengali, 2009) is adopted to solve the problem in this paper.

Generally, the period of H2RTs will be more than 3 years starting from the 1 AU orbit. The minimum period H2RT will be very beneficial in terms of engineering. The trajectory period is related to the sail performance and its perihelion. The minimum period H2RT under such constraints is shown in Fig.4, obtained in a 3D dynamic model. The lightness number of the sail is 0.6 corresponding to 3.558mm/s$^2$ at 1AU. The perihelion is set to be at the same side as the starting point at 0.2AU away from the Sun. The optimal control history of attitude angles and some other variables are shown in Fig.5 and Fig.6.

The direction of the angular momentum of the sailcraft changes twice in a period as seen in Fig.5. The sailcraft has an initial counterclockwise orbital direction and maneuvers away from the Sun to its aphelion. After the vanishment of the orbital angular momentum, the sailcraft moves clockwise to the perihelion point *D* shown in Fig.4. For a given sail, its magnitude of acceleration at 0.2AU will be 25 times greater than that at 1 AU. As shown in Fig.5, the sailcraft can stay for a year near each aphelion while the period of the H2RT is less than 4 years. So the H2RT is suitable for observing the areas near its two aphelia. Another advantage is that the sailcraft will pass its perihelion in a short time. This will be beneficial for the sail design concerning the constraints about the maximum allowed sail film temperature (Rowe et al. 1978). The control history of the sail attitude angles in the first phase is presented in Fig.6. Note that the sail clock angle changes instantaneously with 180° because 360° is the same as 0° for a sail, while the cone angle is continuous. The control history of the attitude angles differs from Vulpetti's results in three aspects. First, the value of the cone angle obtained from the time-optimal control model is varied. Second, the sign of the transverse component of the sail acceleration in a half period is also varied; this was always negative in Vulpetti's interstellar missions. Third, the perihelion velocities of the H2RTs are circular or elliptic speed, in contrast to the hyperbolic speed achieved by Vulpetti. Using such a speed, the sailcraft can form a clockwise elliptic orbit near the Sun with no sail accelerations. It is also different from Mengali's solution (2010) because there are no assumptions about the trajectory. The minimum period H2RT located in the ecliptic plane is reasonable with the suitable constraints.

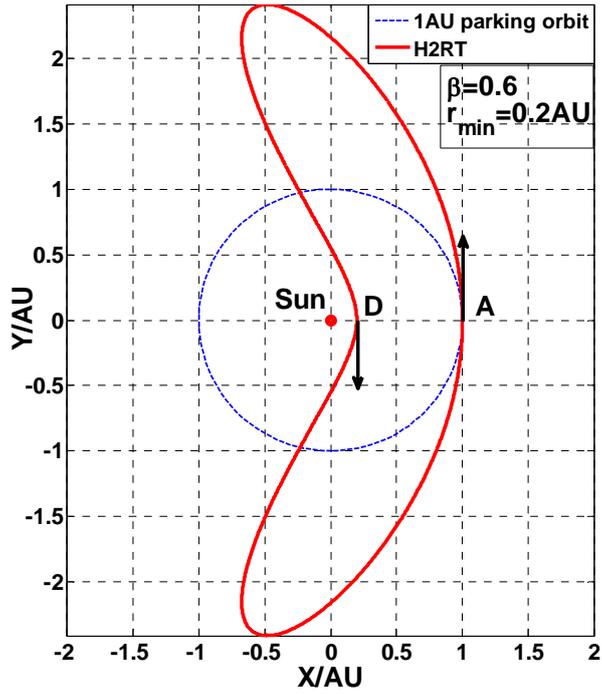

Fig.4   H$^2$-reversal trajectory with given lightness number 0.6

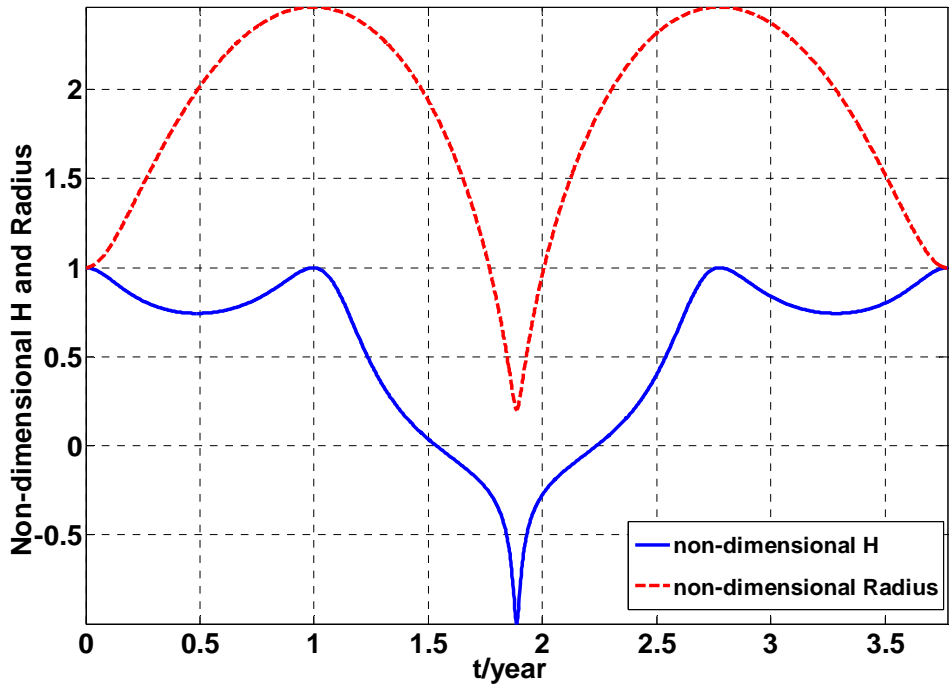

Fig.5   History of non-dimensional sail's H and Radius

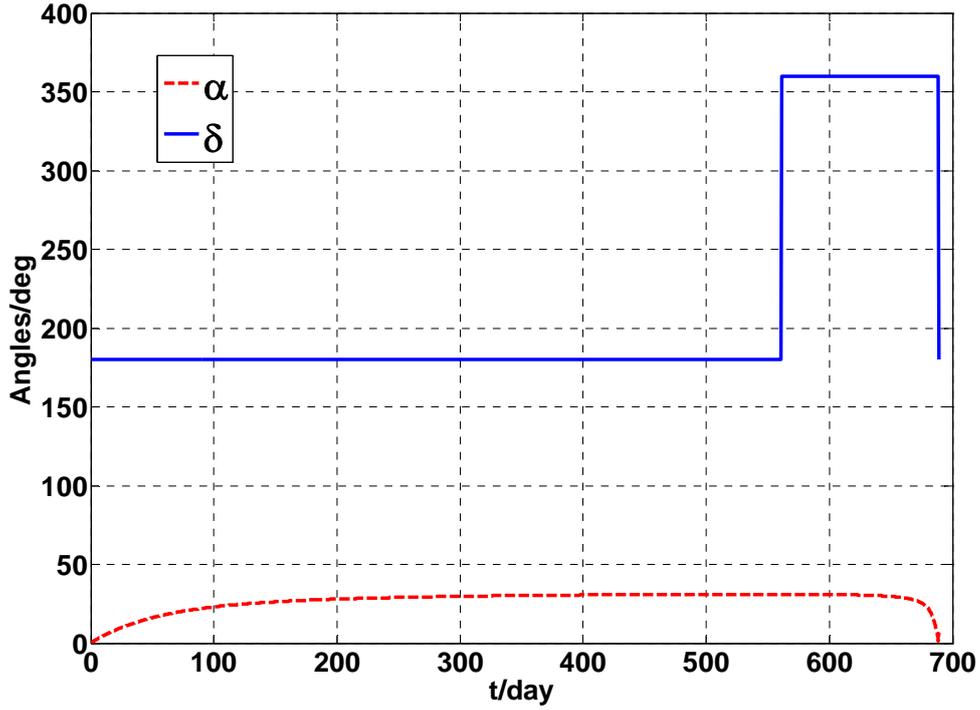

Fig.6　The first phase control history of attitude angles

　　If there is a constraint of position *z* in the trajectory, a 3D H2RT can be obtained, as shown in Fig.7. Taking the lightness number $\beta$=0.7 and *z*=0.1AU as an example, its period is 2.3 years and the trajectory is symmetrical with the plane *xoz*. After the achievement of the above 3D H2RT, it will be easy to find another H2RT with final constraint *z*= -0.1AU. The relationship between the two H2RTs is symmetrical with plane *xoy*. As seen in Fig.8, the variation of the 3D H2RT's control angles is very different from the 2D instantaneous change (remember that 360° is the same as 0° for a sail) in a round trip. Actually, the instantaneous change of the sail clock angle will take place at the starting point and the perihelion, while the cone angle is always continuous. The range of clock angle variation at perihelion is about 10 degrees which can be easily achieved. For a periodic orbit of more than 3 years, the mission time is generally enough during one round trip. The evolution of sail attitude angles as shown in Fig.8 is consistent with the proof of the symmetry property.

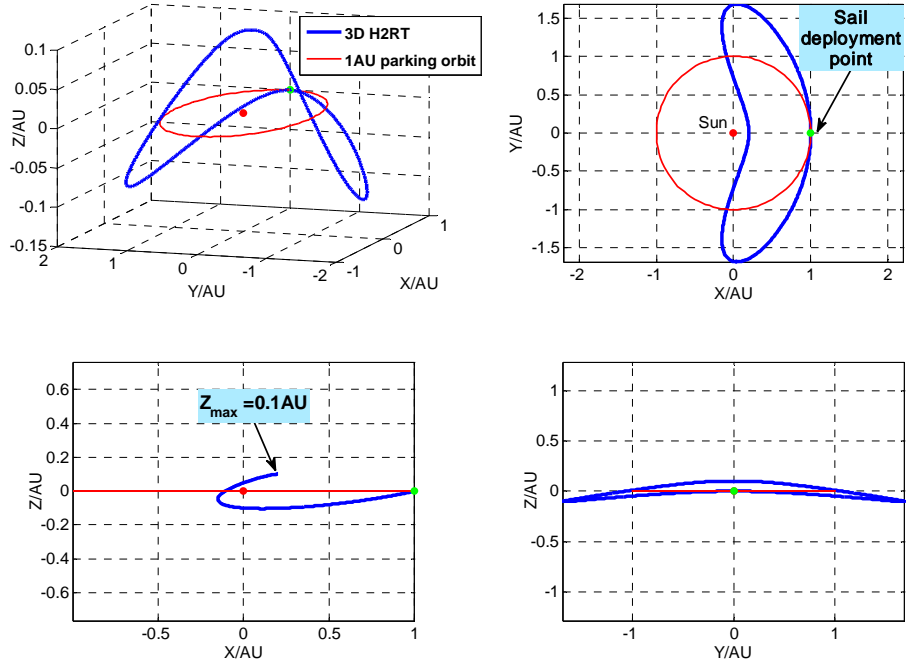

Fig.7  Presentations of the 3D H2-reversal trajectory

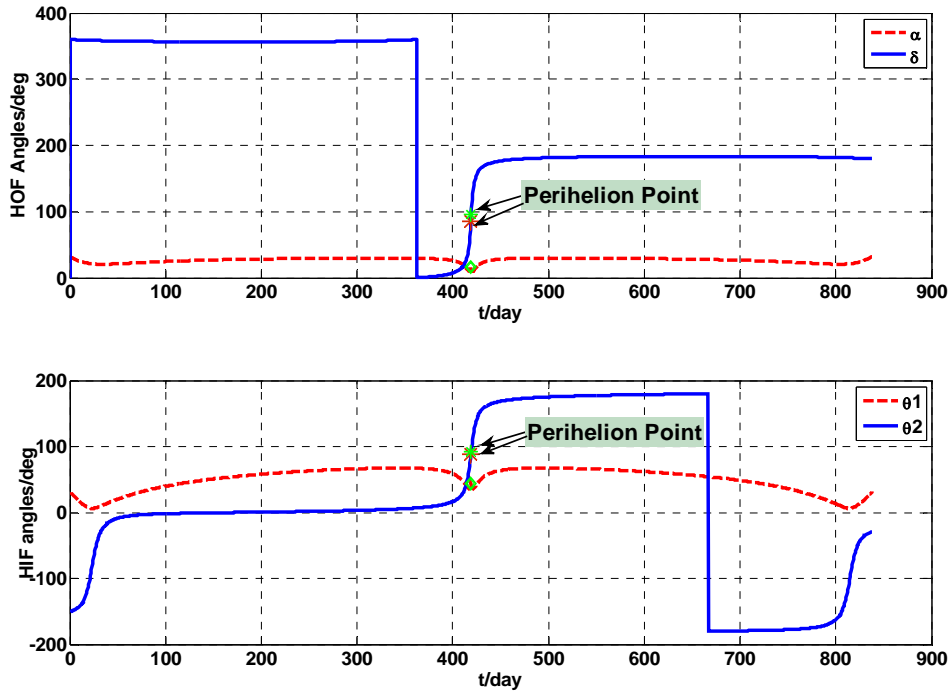

Fig.8  One periodic control history of sail attitude angles

## 3. MISSION APPLICATIONS of H-REVERSAL TRAJECTORY

The H-reversal trajectory has many good properties, and its mission applications are addressed in this section. Some special H-reversal trajectories will be used to complete different space missions. With regard to trajectory design, three main applications are examined in detail.

**3.1 Space Observation**

To enhance our knowledge of the cosmos, a large number of observatories have been launched into orbit. Some of them are designed to survey the entire sky, while others focus on certain parts of the sky. Space environment monitors and orbital telescopes have helped researchers make important discoveries about our universe, ranging from planets and stars to galaxies and cosmological phenomena[3]. Limited to conventional propulsion systems, many space monitors or telescopes have been launched into low-Earth orbits (Aldo 2003). To learn more about the cosmos and the solar system, orbits suitable for in-situ observations will be needed in the near future. A family of trajectories for solar space observation and astronomical observatories are presented in the following section.

Taking the current technology into account the aphelion of the H2RT will be limited to 8AU in the plane *xoy*. The space between Mercury orbit and Saturn orbit can be observed by selecting a different departure time or a different perihelion[4]. Some sample orbits are shown in Fig.9 in the ecliptic plane. There is a special orbit called the 'Self-loop orbit', whose perihelion is 1AU away from the Sun. To avoid a collision with the Earth, the semi-period of the Self-loop orbit should not be an exact multiple of the Earth's period. However, for other H2RTs, their semi-period should be better an exact multiple of the Earth's orbit because this situation can guarantee that the sailcraft and the Earth arrive at the Sun-perihelion line at the same time. This will be better for the data transmission between the sailcraft and the Earth. To extend the observation space, 3D H2RTs will be designed to detect the space not in the ecliptic plane. To illustrate the observation space, 3D H2RTs are shown in Fig.10 with $\beta$=0.7 and $x_{min}$=0.2AU. The figure shows that their projection in plane-*yoz* varies with $z_{max}$ in *z* direction. The variation of the H2RT orbit period is shown in Fig.11 with respect to $z_{max}$. There is a quasi-linear relationship between the orbital period and the constraint $z_{max}$. In Fig.11 the H2RT with a required period can be achieved by choosing the corresponding value of $z_{max}$. For example, if the required period is 4 years, meaning the sailcraft and the Earth will pass through the *xoz* plane at the same time, the value of $z_{max}$ is about 0.63AU right up the ecliptic plane in the *xoz* plane. The sailcraft can stay for about a year near every aphelion about 2.5AU away from the Sun. The minimum distance between the sailcraft and the Earth is about 1.0AU. If such a distance is difficult for data transmission, the Self-loop H2RT can be extended to 3D with a given $z_{max}$ and an exact semi-period. Every time the Earth goes through the starting point (sail deployment point), the sailcraft passes its perihelion right up to the Earth with the distance $z_{max}$.

---

[3] http://en.wikipedia.org/wiki/Space_observatory
[4] Here, the perihelion of H2RTs refers to the flying inward trajectories. For the outward trajectories, it is the crossing point between the trajectory and the symmetric axis.

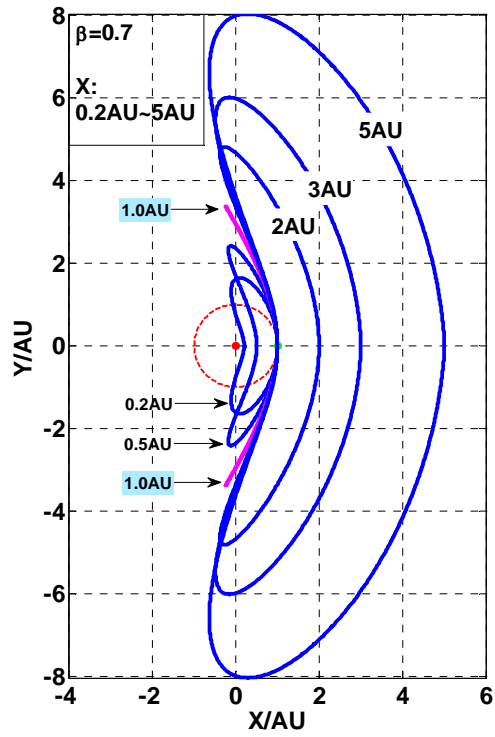

Fig.9  H2RTs in the ecliptic plane

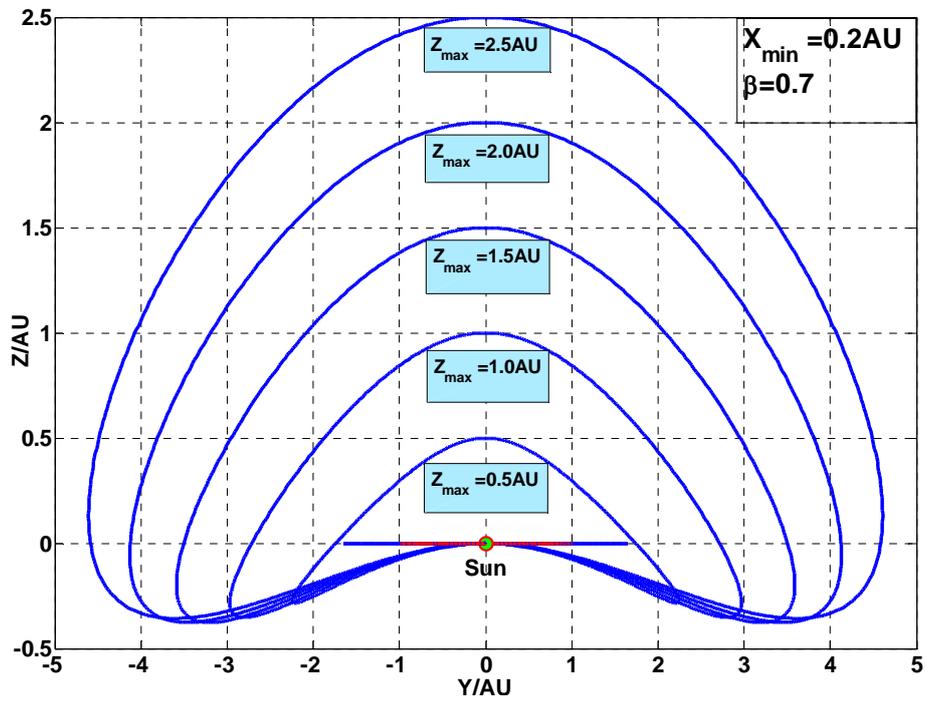

Fig.10  3D H2RT profile in *YOZ* with $X_{min}$=0.2AU

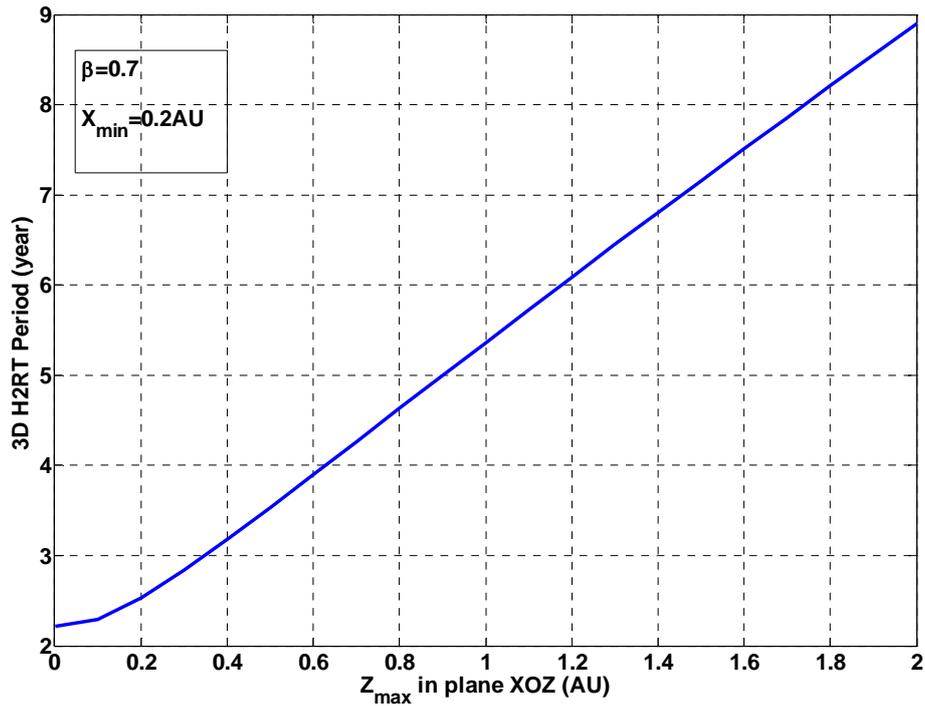

Fig.11    3D H2RT period via $Z_{max}$

**3.2 Heliocentric Transfer Orbit**

Owing to the continuous sail radiation force, the sailcraft is capable of exotic non-Keplerian orbits that are impossible for any other type of spacecraft. One of the Sun-centered non-Keplerian orbits is the heliocentric circular displaced orbit (McInnes 1998). The orbit can be displaced high above the ecliptic by directing a component of the solar radiation pressure force normal to the orbit plane. In particular, the orbit period may be chosen to minimize the required sail lightness number to obtain a given displacement distance. Circular displaced orbits can be used to provide continuous observation of the solar poles or to provide a unique advantage point for infrared astronomy. Another family of Sun-centered orbits is high inclination orbits such as the solar polar orbit. Our current understanding of the Sun and its atmosphere is severely limited by a lack of observation of the Polar Regions. The Solar Polar Imager (SPI) mission places a sailcraft in a 0.48 AU circular orbit around the Sun with an inclination of 75° (Wu et al. 2006; Alexander et al. 2005). The SPI may be able to predic space weather, and it provides an important step in improving our understanding of the physics governing solar variability. In view of the scientific significance of the SPI mission,[5] ESA (European Space Agency), with the assistance of University of Glasgow, has carried out the mission with a moderate solar sail. Both the displaced orbit missions and the SPI mission utilize counterclockwise orbits. Studies of sailcraft moving in clockwise orbits would probably provide some new information about our solar system. However, it is difficult for moderate solar sails, and impossible for conventional propulsion mechanisms, to realize such missions.

In this section, a new mission application of the H-reversal trajectory will be discussed: Heliocentric transfer orbit. When an H2RT reaches its perihelion, if there is no sail acceleration, the sailcraft will form a clockwise orbit. For the ecliptic H2RT obtained in the time-optimal control model, its velocity at its perihelion (Point *D* in Fig.3a) is always circular or elliptic velocity which can be also restricted. If there are no additional requirements regarding the orbit, the sailcraft will orbit the Sun in an elliptic or circular orbit in clockwise direction. The perihelion of the H2RT (Point $D_2$ in Fig.12) will be the perihelion or aphelion of the heliocentric orbit. The 2D H-reversal orbit given in Fig.12 from point *A* to point $D_2$ is an orbit in the ecliptic plane. Such a transfer trajectory can allow for an in-situ observation near the aphelion of the H2RT and then send the sailcraft to the heliocentric orbit.

---

[5] http://sci.esa.int/science-e/www/object/index.cfm?fobjectid=36025

With such a half H2RT, the sailcraft can also be transferred to a high inclination o clockwise orbit. The inclination of the orbit can be given in analytical form

$$i = \tan^{-1}\left(z_{\max}/x_{\min}\right) \quad (15)$$

Considering the temperature limit of the sail, with $z_{\max}$ not less than 0.2AU, $x_{\min}=0$, the sailcraft will be placed into a polar orbit with an inclination of 90º. As mentioned previously, the inclination of the SPI orbit is 75º, so with a given sail lightness number and $x_{\min}$, the value of $z_{\max}$ can be obtained by Eq.(6). Then, the heliocentric H-reversal transfer trajectory can be achieved. The high inclination orbit discussed previously is achieved with no sail force after point $D_3$ (as seen in Fig.12). Actually, when the sailcraft arrives at its perihelion (Point $D_3$), its orbital velocity is along with the opposite direction of axis *oy*. If the value of the velocity is constrained to be the same with the local circular orbit, the family of Sun-centered displaced orbits can be obtained with suitable control laws after point $D_3$. The period of a 0.2AU circular orbit is about 32 days. Within an affordable mission time, the sailcraft can be transferred from the high inclination circular orbit to the Sun-centered displaced orbit with sail force from point $D_3$.

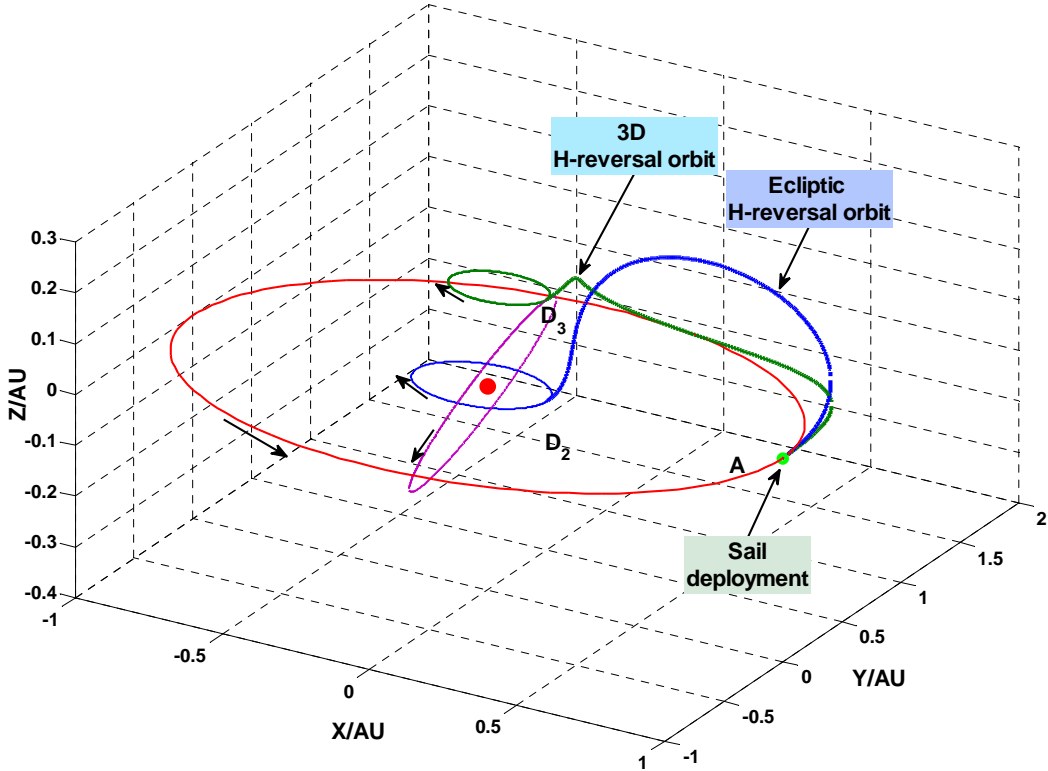

Fig.12    Heliocentric H-reversal transfer orbits

**3.3 Collision with Asteroids near Earth**

Most scientists believe that the threat of asteroids or NEOs (Near Earth Objects) hitting the Earth is real, although they cannot predict when such an event will occur (Slyunyayev et al. 2003; Degtyarev et al. 2009). Astronomers are increasingly concerned about the challenges posed by dangerous asteroids. Effective actions should be taken to prevent the collision of an asteroid with the Earth, as the impact of an asteroid larger than one kilometer in diameter hitting the Earth could cause global devastation. Studies examining how to avoid potential asteroid or other NEO impacts have been conducted in both the United States and Russia. There are three possible methods available once a threat has been identified: Evacuating the Storm, Destruction, and Deflection/Acceleration/Deceleration (Jonathan 2003). Plans have already been announced by ESA to conduct an experiment to see if asteroids can in fact be deflected away from the Earth (Joris 2010). The Destruction strategy can be carried out by the H-reversal trajectory. The sketch map for the collision trajectory is shown in Fig.13. Most orbits concerning a potential impact with an asteroid are similar to the orbit of projection in the ecliptic plane. The asteroid will pass close to Earth at point *B* or *C* in the near future. To avoid such a collision, an H-reversal trajectory can be designed to make the sailcraft hit against the asteroid at point *D*. The main advantage of such a collision is in maximum impact energy because the two objects have opposite velocities at the collision point. The other advantage is that the collision point can be selected by choosing different launch windows. As discussed previously, the $x_{min}$ of the H2RT can be from 0.2 AU to 5 AU or more away from the Sun, so any asteroid detected near the Earth can be impacted by the sailcraft in a pre-designed H2RT. The danger of the Destruction strategy is that incomplete destruction would subject the Earth to multiple impacts from fragments of the original object. In this case, the collision point may be selected at the aphelion of the asteroid, which is generally about 1AU away from the Earth.

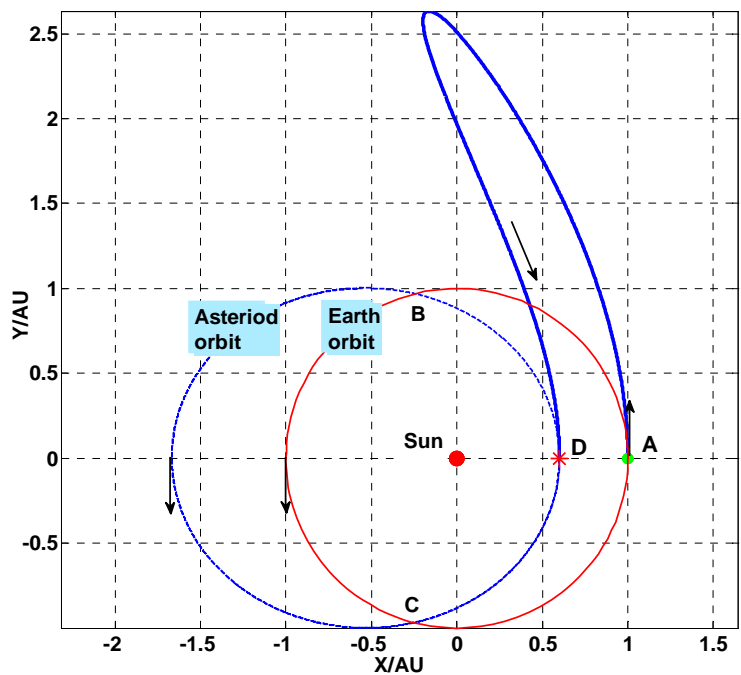

Fig.13    Sketch map for the collision trajectory

The asteroid 99942 Apophis will pass very close to Earth in 2029 (Davis et al. 2006) or in 2036, depending on whether it passes through one of several gravitational keyholes (less than 1km in size) during its 2029 encounter. The Apophis is considered a possible impacter at many subsequent encounters and provides a typical example for the evolution of asteroid orbits. Dachwald (2007) proposed impacting Apophis form a retrograde trajectory with a high relative velocity (75-80km/s) during one of its perihelion passages. A head-on collision of a 150kg impactor (in relative velocity 75km/s) on a $4.67\times10^{10}$kg asteroid yields a pure kinetic impact $\Delta v$ of approximate 0.24mm/s providing a $\Delta x$ about 24km in 1 year (Gehrels, T., 1994). The retrograde orbit would be completed in a cranking phase (Dachwald, B., 2007) to the Sun at 0.2AU in a modest solar sail, which would then rendezvous with the asteroid in direct impact. The orbital elements of Apophis, before and after the 2029 encounter, are listed in Table 1. Taking the asteroid Apophis1 before the 2029 encounter as an example, the trajectory in H-reversal mode was applied to hit the Apophis. The required sail acceleration for such a mission will be greater than that required by a cranking orbit while the solar sail steering law will be easier to be achieved. As seen in Fig.14, the asteroid Apophis will impact or fly by the Earth at point $P$. The most effective collision point is its perihelion at 0.746AU. According to the period of Apophis, the launch windows of the sailcraft from a 1AU orbit can be determined easily. Two points were selected to be the collision points located at $P_1$ and $P_2$. Point $P_1$ is the crossing point between the Apophis orbit and the ecliptic H-reversal orbit. Point $P_2$ is the crossing point between the Apophis orbit and a 3D H-reversal orbit. The simulation results are listed in Table 2. The mean relative speed between the two bodies is about 70km/s, which is consistent with Dachwald's results. The two periodic H2RTs can be seen in Fig.14. Both of the two trajectories are obtained in a time-optimal control framework. Therefore, the collision velocity is not at its maximum and there is an intersection angle between the collision velocities. Additionally, the high crash energy is not from the high performance solar sail but from the head-on collision sailcraft in an H-reversal trajectory. To obtain the largest head-on impact velocity, an energy-gain optimal control model should be applied to complete the collision mission. In this way, the collision efficiency will increase. However, this case will not be discussed here. Based on the previous discussion and mission design, if there is a need for a collision in the near future, the Destruction strategy using the H-reversal trajectory would be the preferred option.

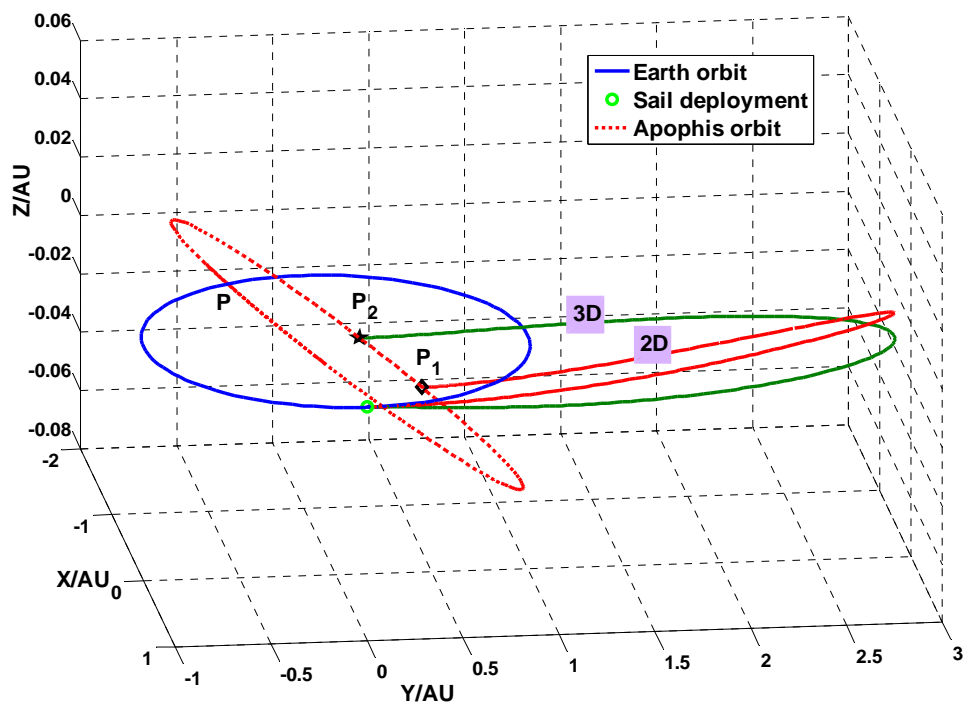

Fig. 14 Collision trajectories in 2D/3D H-reversal mode

Table 1 Orbital elements of the Earth-impacting Apophis orbit variations Ap1 and Ap2[6]

|  | Before 2029-encounter | | After 2029-encounter | |
|---|---|---|---|---|
|  | Apophis1 | Apophis2 | Apophis1 | Apophis2 |
| MJD2000 | 1914.5 | 1914.5 | 13154.5 | 13154.5 |
| $a$ [AU] | 0.9223913 | 0.9223912 | 1.1082428 | 1.1082581 |
| $e$ | 0.191038 | 0.191038 | 0.190763 | 0.190753 |
| $i$ [deg] | 3.331 | 3.331 | 2.166 | 2.169 |
| $\Omega$ [deg] | 126.384 | 126.383 | 70.230 | 70.227 |
| $\omega$ [deg] | 204.472 | 204.472 | 203.523 | 203.523 |
| $M$ [deg] | 203.974 | 203.974 | 227.857 | 227.854 |

Table 2 Collision scenarios of Apophis with the H-reversal trajectories

| Collision position | Collision date [MJD2000] | Inertial position[AU] | | | Relative velocity[km/s] |
|---|---|---|---|---|---|
|  |  | $x$ | $y$ | $z$ |  |
| $P_1$ | 7675.1 (2021-Jan-05) | 0.761 | 0.0 | 0.018 | 68.83 |
| $P_2$ | 7719.1 (2021-Feb-18) | 0.726 | 0.331 | 0.0 | 71.06 |

## 4. CONCLUSIONS

In this paper, three new applications of the H-reversal trajectory using solar sails in a two-body frame have been presented. The double angular momentum reversal trajectory (H2RT) was originally obtained using a 3D dynamic model with an ideal solar sail. The properties of the H2RT have been discussed in detail for both 2D and 3D trajectories. The minimum period H2RTs were studied in a time-optimal control framework under various constraints. Because the special shape of the H2RTs guarantees a quasi-heliostationary condition near its two symmetrical aphelion points ranging from Earth's orbit to Saturn's orbit, the H2RTs are suitable for in-situ observation trajectories. Another important application of the H-reversal trajectory is transferring the sailcraft to a heliocentric clockwise orbit. The final Sun-centered orbit can be achieved easily in the ecliptic plane, with a high inclination or displacement above or below the Sun. The inclination of the heliocentric orbit can be designed with a simple equation. Sample orbits of each application have been given for clarity. Finally, the H-reversal trajectory was used to simulate a collision with an asteroid passing near the Earth. It was determined that there would be high crash energy in a head-on impact. The collision point between the sailcraft and the threatening asteroid can be selected by choosing different launch windows. It should be emphasized that the lightness number of the illustration orbits given in this paper is not less than 0.5. However, this may not be the minimum value that can produce the H-reversal trajectory in an optimal control framework. This problem of the minimum lightness number and the new shape of 3D H2RTs will be studied in future research.

---

[6] Also available at http://ssd.jpl.nasa.gov/sbdb.cgi?sstr=99942%20Apophis;orb=0;cov=0;log=0;cad=0#discovery

**Acknowledgments**

This work was supported by the National Natural Science Foundation of China (Grants No.10902056 and No. 10832004). Acknowledge Prof. Christopher, Tsinghua University, for meticulous revision of this paper.